\documentclass{ifacconf}
\usepackage{enumitem, xcolor}
\usepackage{url}
\usepackage{graphicx}  
\usepackage{amsmath, amsfonts, amssymb}
\usepackage{natbib}     

\begin{document}
\begin{frontmatter}

\title{Data-Driven Nonlinear State Observation using Video Measurements\thanksref{footnoteinfo}} 

\thanks[footnoteinfo]{This work is supported by the faculty start-up fund from NC State University.}
\author{Cormak Weeks, Wentao Tang} 
\address{Department of Chemical and Biomolecular Engineering, North Carolina State University, Raleigh, NC 27695-7905, USA}

\begin{abstract}              
% Why doing this? 
State observation is necessary for feedback control but often challenging for nonlinear systems. While Kazantzis-Kravaris/Luenberger (KKL) observer gives a generic design, its model-based numerical solution is difficult. 
% What is the technical approach? 
In this paper, we propose a \emph{simple method to determine a data-driven KKL observer}, namely to (i) transform the measured output signals by a linear time-invariant dynamics, and (ii) reduce the dimensionality to principal components. 
% What is the use? Example 
This approach is especially suitable for systems with rich measurements and low-dimensional state space, for example, when \emph{videos} can be obtained in real time. We present an application to a video of the well-known Belousov-Zhabotinsky (B-Z) reaction system with severe nonlinearity, where the data-driven KKL observer recovers an oscillatory state orbit with slow damping. 
\end{abstract}

\begin{keyword}
State observation, Nonlinear systems, Principal component analysis, Video data
\end{keyword}

\end{frontmatter}
%===============================================================================

\section{Introduction}
\par Characterizing the nonlinearity in process dynamics is of critical importance to process control. While model-based control methods, especially model predictive control (MPC) \citep{rawlings2017model}, have been well established in the existing literature, the difficulty of obtaining accurate nonlinear models has motivated the development in the directions of both nonlinear system identification \citep{nelles2020nonlinear}, especially for neural models \citep{ren2022tutorial}, and model-free data-driven nonlinear control \citep{hou2013model, tang2022data}. 
This work focuses on \emph{model-free} control, and is motivated by as well as complements our previous works on input--output data-driven control methods \citep{tang2019dissipativity, tang2021dissipativity, tang2023dissipativity}, where machine learning techniques are adopted to analyze input and output trajectories for the synthesis of output-feedback controllers. 

\par For nonlinear process control, state-space forms are the \textit{de facto} standard representation -- when a state-space model is available, an optimal or pole-placing state-feedback controller is synthesized, and as a prerequisite for such state feedback, a state observer should be first designed to estimate the states from input and output history \citep{sontag1998mathematical, haddad2008nonlinear}. 
A \emph{state observer} refers to an auxiliary dynamical system that takes the plant inputs and outputs as its inputs, and outputs a signal as the estimation of the states \citep{kravaris_advances_2013}. For observable linear systems, the Luenberger observer \citep{luenberger_observers_1966} achieves exponential decay in the state observation errors. For nonlinear systems, observer designs based on extended Kalman filters (EKF), input--output linearization, high-gain feedback, and other ideas have been proposed; a comprehensive  recent review can be found in \cite{bernard2022observer}. 

\par Here we in particular focus on a special type of observers, initially proposed in \cite{kazantzis1998nonlinear} as an extension of the Luenberger observer in nonlinear systems with a local exponential convergence guarantee. Their global existence and properties were studied in \cite{andrieu_existence_2006}, which also pinned down its naming as \emph{Kazantzis-Kravaris/Luenberger (KKL) observers}. Multiple extensions of KKL observers were proposed in the literature \citep{duan2020nonlinear, kravaris2021functional}. 
We especially note that, without the knowledge of a state-space model, KKL observer provides a salient structure (i.e., a linear state dynamics and a nonlinear static output map) that is \emph{amenable to machine learning}, especially neural network approaches \citep{niazi2023learning, miao2023learning, tang2023synthesis}. 
To avoid the nonconvexity in neural network training, \cite{tang2023data} recently proposed a linear parameterization of the observer in the form of Chen-Fliess series, which can be recursively learned through online least squares.  

\par In this paper, we consider the scenarios where a \emph{rich} amount of measured variables are obtained from an (autonomous) system with \emph{low-dimensional} state space. This is motivated by the growing interest in utilizing image/video information for process modeling, monitoring, and control \citep{shi2020real, jiang2021convolutional, wang2022predictive}, which in principle provides higher capacity of understanding and handling the nonlinear process dynamics \citep{chiang2022towards}. 
However, \emph{exploiting such rich visual information for state observation}, despite being a natural usage, has not been reported so far to our knowledge. The present work thus aims to bridge this gap, especially for reactive systems with severe nonlinearity. To showcase the applicability, we will consider as case study a representative nonlinear autonomous system -- the Belousov-Zhabotinsky (B-Z) reaction system \citep{zhabotinsky_history_1991}, which exhibits repetitive oscillations across a spectrum of colors due to the periodic variations of the concentrations of reactive species. 

\par We argue that when video measurements are available, the output mapping of the KKL observer can be approximated as an \emph{affine} one, thus allowing a simple approach for data-driven KKL observer synthesis. 
Specifically, we propose to determine the observer through two steps, namely (i) passing the output measurements through a pre-assigned linear time-invariant (LTI) dynamics, to obtain the observer states that stand for a lifting of the process states, and (ii) reducing the dimensionality of the observer states through principal component analysis (PCA) that retains the information in a low-dimensional subspace with most data variations. Through these two steps, we obtain as outcomes a state trajectory that is \emph{topologically equivalent} (in the sense of a diffeomorphism) to the actual one. 
When the true states (with physical meanings) are not known, recovering such an equivalent representation is sufficient for subsequent state-space modeling, dynamical analysis, or potentially, model-free state-feedback control. 

\par The remainder of the paper is organized as follows. We provide preliminaries of the KKL observer in Section \ref{sec:KKL} and discuss the PCA algorithm and its applicability in Section \ref{sec:PCA}. Then, focusing on an oscillatory reaction system with video measurements, we apply the proposed PCA-based data-driven KKL observer approach in Section \ref{sec:BZ}, showcasing its capability of recovering the limit cycle behavior. Conclusions are given in Section \ref{sec:final}.

\section{KKL Observer for Nonlinear Systems}\label{sec:KKL}
\par Consider a nonlinear autonomous system in continuous time, described by the equations: 
\begin{equation}
\begin{aligned}
    \dot{x}(t) &= F(x(t)), \\
    y(t) &= H(x(t)),
\end{aligned}
\end{equation}
where \(x \in \Rset^n\) is the vector of state variables, and \(y \in \Rset^p\) represents the measured outputs. It is assumed that \(F: \Rset^n \rightarrow \Rset^n\) and \(H: \Rset^n \rightarrow \Rset^p\) are supposed to be smooth to guarantee the existence and uniqueness of the flow. However, the algebraic expressions of $F$ and $H$ are unavailable for model-based state observation. 
To pursue a data-driven state observer for this black-box dynamics, we assume that the state space is intrinsically low-dimensional (i.e., $n$ is small) and that the available measurement is rich, with a sufficiently high output dimension (\(p\)). 
The low-dimensional state space typically arises from process units with dissipative flows. In such systems, the dynamics naturally separates into a fast self-stabilizing subsystem that converges to an inertial manifold and slow patterns that dominate the long-term behavior. See, e.g., \cite{christofides2001control, baldea2012dynamics}.
The high-dimensional measurement typically comes from a video captured from the system, when \(p\) in principle can be as large as the number of pixels in each image frame. 

\par The most classical form of state observer is known as the Luenberger observer \citep{luenberger_observers_1966}, which is suitable for linear systems (i.e., \(F(x) = Fx\) and \(H(x) = Hx\) for matrices \(F\) and \(H\)) and represented as:
\begin{equation}
\begin{aligned}
\dot{z}(t) &= Az(t) + By(t), \\
\hat{x}(t) &= T^{\dagger}z(t).     
\end{aligned}
\end{equation}
Here $z \in \Rset^{n_z}$ (\(n_z\geq n\)) is the observer states, and $\hat{x}$ is the vector of state estimates as the observer outputs. The observer dynamics is a linear one specified by the matrices of appropriate dimensions (\(A\), \(B\), and \(T^{\dagger}\)). 
To ensure that the observation error \(e := \hat{x}(t) - x(t) \rightarrow 0\) as \(t \rightarrow \infty\), the following three steps are taken: 
\begin{enumerate}[label=(\roman*)]
    \item Assigning the observer state dynamics \((A, B)\), under the restriction that \((A, B)\) must be a controllable pair and \(A\) should be Hurwitz; 
    \item Solving the Sylvester equation 
    \begin{equation}
        TF - AT = BH
    \end{equation}
    for \(T \in \Rset^{n_z\times n}\) under the given \((A, B)\); and
    \item Finding \(T^{\dagger}\) as a left pseudo-inverse of \(T\). 
\end{enumerate} 

\begin{rem}
    A special case occurs when \((A, B)\) is chosen such that \(A = F - BH\), resulting in \(T = I\). The observer then reduces to a linear dynamics driven by the ``innovation'' quantity $y - H\hat{x}$: 
    \begin{equation}
     \dot{\hat{x}}(t) = F\hat{x}(t) + B(y(t) - H\hat{x}(t)). 
    \end{equation}
    The observer gain \(B\) can be optimally designed via solving an algebraic Riccati equation. This is the (continuous-time) Kalman filter. 
\end{rem}

For nonlinear systems, the Luenberger observer was generalized to the concept of Kazantzis-Kravaris/Luenberger (KKL) observer. The KKL observer is expressed as:
\begin{equation}
\begin{aligned}
    \dot{z}(t) &= Az(t) + By(t), \\ 
    \hat{x}(t) &= T^{\dagger}(z(t)),
\end{aligned}
\end{equation}
where the observer states \(z\) are still driven by the process outputs $y$ through an LTI dynamics \((A, B)\), but the output map is replaced by a nonlinear one, $T^\dagger(\cdot)$, which is the left-pseudoinverse of a smooth nonlinear mapping $T: \Rset^n \rightarrow \Rset^{n_z}$ that is locally non-degenerate everywhere (i.e., $T$ is an immersion). Here, by left-pseudoinverse, we mean that \(T^\dagger\) is a mapping satisfying \(T^{\dagger}(T(x)) = x\) for all \(x\) on the state space, or simply \(T^{\dagger} \circ T = \mathrm{id}\). 
It can be verified that, if $T$ is the solution to the following system of partial differential equations (PDEs): 
\begin{equation}\label{eq:PDE}
    \frac{\partial T}{\partial x}(x)F(x) = AT(x) + BH(x), 
\end{equation}
then \(z\) becomes an asymptotic estimate of \(T(x)\), representing the transformed states. Here $\partial T/\partial x$ refers to the Jacobian matrix of $T(x)$. The mapping \(T^\dagger\) transforms the observer states \(z\) back into the estimated state variables \(\hat{x}\) in \(\mathbb{R}^n\), thus generating asymtotic state estimates (due to the non-degeneracy of \(T\)). 

\par The existence of KKL observers for nonlinear systems was proved in the literature under mild back-distinguishability assumptions on the flow \citep{andrieu_existence_2006, brivadis2023further}. We note that for nonlinear systems, the order of state observer dynamics is not guaranteed to be equal to \(n\), with \(n_z > n\) expected. 
Specifically, if the pole placements in \(A\) are allowed to be complex, the observer order should be chosen as \(n+1\) for a single-output system, which implies that with \(p\) outputs, a sufficient order of state observer is \(n_z = p(n+1)\). This makes \(T\) a \emph{lifting} into a higher-dimensional space and thus \(T^\dagger\) a \emph{dimensionality reduction} from \(p(n+1)\) to \(n\) dimensions. 

\par While the construction of this inverse mapping is crucial for observing the original state variables, its reliance on the a model-based PDE system often makes it not amenable to numerical solution. 
Nevertheless, it was noted \citep{tang2023data} that even if the states (\(x\)-data) are unavailable for the supervised learning of the observer, \emph{dimensionality reduction} techniques on the \(z\)-data can determine a mapping \(P: \Rset^{n_z} \rightarrow \Rset^n\), which generally makes $P\circ T$ a \emph{diffeomorphism} (namely a smooth bijective mapping that has a smooth inverse).
That is, $P(z)$ will yield a topologically equivalent representation of the states, containing all the state information despite the absence of the ``ground truth''. For dimensionality reduction, there are abundant tools of machine learning. Next, we introduce and justify the use of PCA algorithm to this end.

\section{Dimensionality Reduction for State Observation using PCA}\label{sec:PCA}
In a dimensionality reduction algorithm, the reduction mapping \(P\) is learned (i.e., optimized according to a sample) within a class of candidate mappings, denoted as \(\mathcal{P}\). In order that \(P\circ T\) becomes a bijection through such a learning procedure, we make the following assumption.
\begin{assum}\label{assum:ground}
    The ``ground truth'' mapping to reconstruct the states, \( T^\dagger \), can be left-composed by an (unknown) bijection \(S: \Rset^n \rightarrow \Rset^n\), and then be contained in the model class \( \mathcal{P} \), i.e., \(S\circ T^\dagger \in \mathcal{P}\). 
\end{assum}

This assumption in practice is satisfied only approximately. To ensure better approximation, \(\mathcal{P}\) should be a rich model class. For example, in \cite{tang2023data}, we proposed to use kernel PCA in the presence of low-dimensional measurements and large data, so that the \(\mathcal{P}\) can cover the linear span of a large number of kernel functions. 
In a different vein, in this paper, we consider the setting that the measurement dimension \(p\) is large, i.e., the system's output mapping \(H = (H_1, \dots, H_p)\) contains a large number of nonlinear functions. As such, for the underlying immersion \(T\) as a very high-dimensional vector-valued mapping, its left-pseudoinverse \(T^\dagger\) not only exists but also has a sufficiently large degree-of-freedom. 

\par Specifically, we will consider the case that the model class \(\mathcal{P}\) contains all affine mappings from \(\Rset^{n_z}\) to \(\Rset^n\). That is, 
\begin{equation}
    \mathcal{P} = \{ P : z \mapsto Qz - q, \enskip Q \in \Rset^{n\times n_z}, \, q \in \Rset^n\}.    
\end{equation}
With Assumption \ref{assum:ground}, we assume that there exists some ``true'' model \(P^\ast(z) = Q^\ast z - q^\ast\), such that for some bijection \(S^\ast\) and for some pseudo-inverse \(T^\dagger\), we have \(S^\ast \circ T^\dagger = P^\ast\). 
This justifies the use of PCA \citep{jolliffe_principal_2016} as an appropriate tool, which determines a dimensionality reduction mapping in an affine form: 
\begin{equation}\label{eq:affine}
    \hat{x} = P(z) = Qz - q.  
\end{equation}

\par Now, we summarize the major steps in the PCA approach. 
First, by passing the measured output signal of the system \(y(\cdot)\) through an LTI dynamics \((A, B)\), the observer states \(z(\cdot)\) can be simulated. With a certain sampling interval, we may store the data of \(z[k]\) in discrete time \(k=1,2,\dots,N\) into a matrix \(Z\) is of shape \(N\times n_z\), namely 
\begin{equation}
    Z = 
    \begin{bmatrix}
        z[1]^\top \\ z[2]^\top \\ \vdots \\ z[N]^\top
    \end{bmatrix}.
\end{equation}
With the data being a video captured from the physical process, the number of discretized time instants \(N\) is the number of frames from the video, and the number of data features is $n_z$. The first step in PCA is to determine the average and sample covariance of each component:
\begin{equation}
    \bar{z}_j = \frac{1}{N}\sum_{k=1}^N z_j[k], \quad 
    s_j^2 = \frac{1}{N-1}\sum_{k=1}^N \left( z_j[k] - \bar{z}_j \right)^2, 
\end{equation}
and thereby calculate the whitened data matrix \(\tilde{Z}\), whose \((k, j)\)-th entry is:
\begin{equation}
    \tilde{z}_j[k] = \frac{z_j[k] - \bar{z}_j}{s_j}, \quad j=1,\dots,n_z, \enskip k=1,\dots,N. 
\end{equation}

\par Then, we compute the covariance matrix $\Sigma$:
\begin{equation}
\Sigma = \frac{1}{N-1} \tilde{Z}^\top \tilde{Z}, 
\end{equation}
and perform eigenvalue decomposition on the covariance matrix $\Sigma$: 
\begin{equation}
    \Sigma = U\Lambda U^\top 
\end{equation}
where $U = [u_1, \dots, u_{n_z}]$ is the orthogonal matrix comprising of unit eigenvectors of $\Sigma$ as its columns and $\Lambda = \mathrm{diag}(\lambda_1, \dots, \lambda_{n_z})$ is the diagonal matrix of the corresponding eigenvalues, which capture the variations of data along the directions of the eigenvectors. It actually suffices to perform a singular value decomposition algorithm on the data matrix $\tilde{Z}$, which avoids the explicit calculation of \(\Sigma\). 
From the PCA result, the \(i\)-th principal component of any sample point \(z[k]\) is \(u_i^\top \tilde{z}[k]\). Hence, the principal components of all data points can be stored as \(\tilde{Z}U\). 

\par Then we reduce the dimensionality from \(n_z\) to \(n\). To maximize the variation encompassed by the reduced set of principal components, we select the \(n\) leading eigenvalues and their corresponding eigenvectors. Assuming that the diagonal of \(\Lambda\) has been sorted in the descending order, we denote by \(\tilde{U} \in \Rset^{n_z \times n}\) the matrix formed by the first \(n\) columns of \(U\). The proportion of total variance retained in the \(n\)-dimensional subspace of these principal components is \( \sum_{i=1}^n \lambda_i / n_z\), where $\lambda_i$ represents the \(i\)-th eigenvalue. Then, the principal components of any sample point \(z\) form a vector
\begin{equation}
    P(z) = \tilde{U}^\top\tilde{z} = \tilde{U}^\top S^{-1} (z - \bar{z}), 
\end{equation}
where \(S = \mathrm{diag}(s_1, \dots, s_{n_z})\). Let \(Q = \tilde{U}^\top S^{-1}\) and \(q = Q\bar{z}\). We have determined a dimensionality reduction mapping in the form of \eqref{eq:affine}. 

\begin{rem}
    The PCA algorithm here introduced is an \emph{offline} one, which computes the reduction mapping after a sufficiently large sample is collected, and this mapping is not updated with new data. An online version, known as recursive PCA, was proposed based on the perturbation of spectral decomposition and used for online process monitoring in \cite{li2000recursive}. An online kernel PCA algorithm can be found in \cite{honeine2011online}. 
\end{rem}

\begin{rem}
    The complexity theory of subspace learning has established general probabilistic relations between the sample size \(N\), error \(\epsilon\) in the distance between the learned subspace \(\Sigma\) and true subspace \(\Sigma^\ast\), and the confidence level \(1 - \delta\), in the form of 
    \begin{equation}
        \mathbb{P}[\mathrm{dist}(\Sigma^\ast, \Sigma) \leq \epsilon(n, \delta)] \geq 1-\delta. 
    \end{equation}
    See, e.g., \cite{rudi2013sample}, where it was established that the error is such that
    \begin{equation}
        \epsilon(n, \delta) \sim O\left( \left( \frac{\log(n/\delta)}{n}\right)^{-(1-r)/2r} \right), 
    \end{equation}
    where \(r\) is the order of decay for the eigenvalues in descending order. Thus, under Assumption \ref{assum:ground}, it can be claimed that for the actual bijection from the states to the principal components, \(S^\ast\), the expected observation error, defined as 
    \begin{equation}
        e(x) := S^\ast(x) - P\circ T(x), 
    \end{equation}
    vanishes with increasing sample size. Since the Assumption \ref{assum:ground} holds only approximately, the observation error is practically bounded instead. 
\end{rem}

\section{Case Study: Continuously Stirred Oscillatory Reaction}\label{sec:BZ}
\par Now we consider the case study on B-Z reactions. The reaction kinetics is accurately described by the Field-K{\H{o}}r{\"{o}}s-Noyes (FKN) mechanism which is further simplified by the Oreganator model \citep{field_science_2022}. By using video processing techniques, the oscillatory nature of the Oregonator model can be confirmed and the dynamical aspects was investigated in the recent work of \cite{barzykina_chemistry_2020}. 
Motivated by the abundance of videos on the B-Z reaction system due to its use as a tutorial example, in this work, we choose this system as a prototype reactor unit with low-dimensional state space and high-dimensional measurements, where the proposed data-driven state observation approach is applicable. 

\par
\begin{figure}[ht]
\begin{center}
\includegraphics[width=8.4cm]{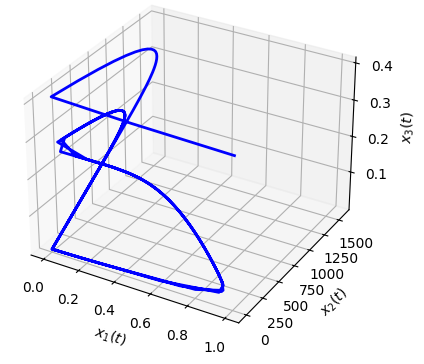}    % The printed column width is 8.4 cm.
\caption{Simulated state trajectory in an Oregonator model for the B-Z reactions. } 
\label{fig:Oregonator}
\end{center}
\end{figure}

\begin{figure}[ht]
\begin{center}
\includegraphics[width=8.4cm]{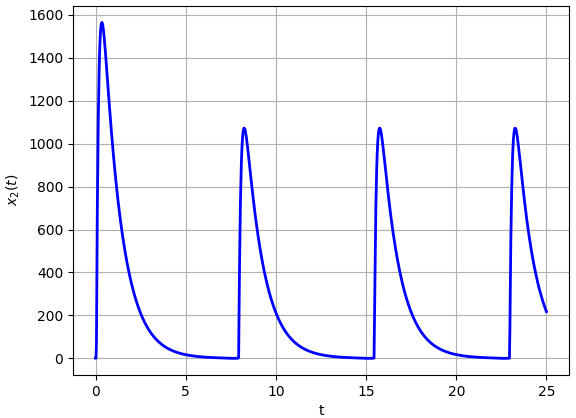}    % The printed column width is 8.4 cm.
\caption{Trajectory of the second state component in the Oregonator model.} 
\label{fig:Oregonator_x2}
\end{center}
\end{figure}

 The Oregonator model for the B-Z reactions reduces the five coupled reaction rate expressions into three differential equations. These three equations can be put as a dimensionless form: 
\begin{equation}
\begin{aligned}
    \epsilon \frac{dx_1}{dt} &= qx_2-x_1x_2+x_1(1-x_1), \\
    \delta \frac{dx_2}{dt} &= -qx_2-x_1x_2+fx_3, \\
    \frac{dx_3}{dt} &= x_1-x_3. 
\end{aligned}
\end{equation}
The constants are related to the rate constants of the coupled reactions (for a detailed explanation, see, e.g., \cite{barzykina_chemistry_2020}). Here the constant values are $\epsilon=3.6\times10^{-2}$, $\delta = 1.2\times10^{-4}$, $f=1$, and $q=2.4\times10^{-4}$. 
Simulating the model from a randomly selected initial condition, we plot the trajectory of the states in a three-dimensional space as in Fig. \ref{fig:Oregonator}. After a short time duration, the trajectory settles down on a limit cycle that has a bow-tie shape seen from the chosen view. In addition, Fig. \ref{fig:Oregonator_x2} plots the trajectory of $x_2$ versus increasing time, which confirms that the orbit is periodic with self-sustained oscillations. 
Now that we assume that the true model is unavailable (and indeed the above model is only approximate), we examine whether the proposed data-driven state observer is capable of recovering this limit cycle behavior. 

\begin{figure}[ht]
\begin{center}
\includegraphics[width=8.4cm]{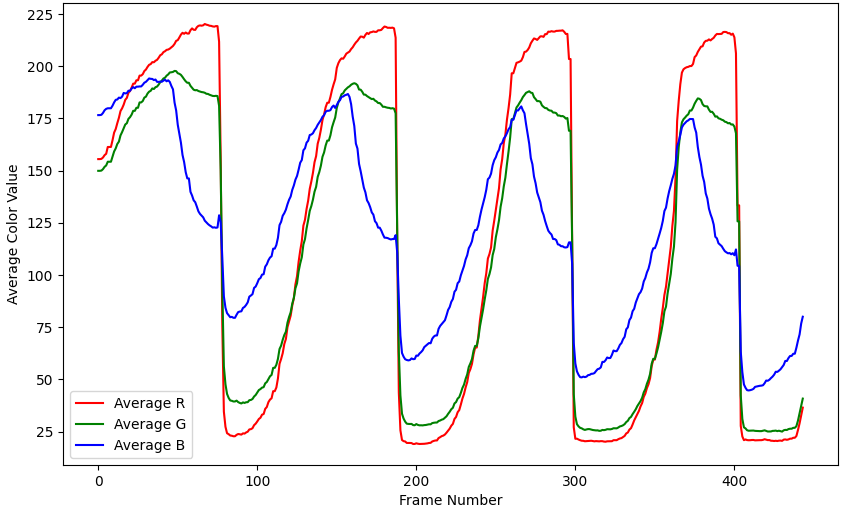}    % The printed column width is 8.4 cm.
\caption{Variation of pixel-averaged color in three channels with increasing frame number.} 
\label{fig:Pixels}
\end{center}
\end{figure}

To this end, we obtained a video from Youtube, showing the reaction occurring in a flask caught by a camera at a fixed position\footnote{The video was uploaded by Bob Kulawiec at Edmund Burke School on February 8, 2023 at \url{https://www.youtube.com/watch?v=ieh9qIkkMJQ}.}. Each video frame has \(360\times640\) pixels, in which a \(150\times 130\)-pixel area approximately covers the flask. Since the solution in the flask is continuously stirred by a magnetic agitator, the solution color is not well captured in the regions with vortices and bubbles. To guarantee data quality, we selected a spatial domain with \(10\times10\) pixels. The total time duration used in this study is \(34\) seconds, which is recorded at a rate of \(0.075\) seconds per frame. 
In each video frame, every pixel is represented by three scalars corresponding to three color channels -- red (R), green (G), and blue (B), respectively. The variation in the RGB color vector at each pixel reflects the changing compositions of the chemical species in the reacton. Thus, each frame is converted to a \(300\)-dimensional vector, i.e., the output dimension of the underlying system is \(p=300\). In Fig. \ref{fig:Pixels}, we plot the variation of the color components (averaged on the \(100\) pixels for each of the three channels) with increasing time. Clearly, a synchronized periodic oscillation exists on all color channels, which entails a limit cycle behavior of the underlying states. 
All the processing steps here utilizes the OpenCV  library (version 4.8.1) in Python on Google Colaboratory. 

\par For simplicity, we assume that the state dimension \(n=3\) is a priori known. Hence, the KKL observer should have a minimal order of \((n+1)p = 1200\). We thus independently assign random numbers \(a_1, \dots, a_{1200}\) within the range of \([1, 50]\) and let \(A = -\mathrm{diag}(a_1, \dots, a_{1200})\) so as to make the observer's LTI dynamics have time scales between \(0.02\) and \(1\) and is also stable. The matrix \(B\) is randomly picked as a full matrix of shape \(1200\times 300\). In general, \((A, B)\) should be a controllable pair, allowing the formulation of a high-dimensional observer state vector \(z\) that encapsulates the nonlinear behaviors of the true states \(x\). 
As pointed out in the previous section, it is reasonable to postulate that the variations within the affine combinations of the observer state components \(z_1,\dots,z_{1200}\) are sufficient to reflect the behavior of the underlying states. 
Hence, PCA is performed to extract the \(3\) principal components, capturing the most data variations, from a superficial \(300\)-dimensional measurement. Such a dimensionality reduction is naturally expected, since the color variation across the pixels should be small. 

\begin{figure}[!t]
\begin{center}
\includegraphics[width=0.85\columnwidth]{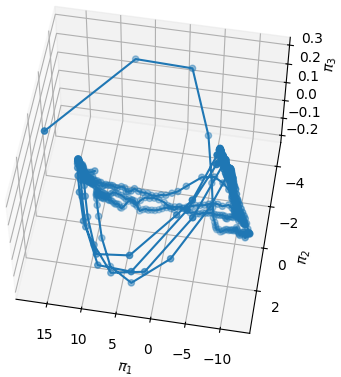}   
\caption{Trajectory of principal components, which exhibits a limit cycle topologically similar to the one predicted by the model.}
\label{fig:KKL}
\end{center}
\end{figure}

\par Fig. \ref{fig:KKL} visualizes the trajectory of the three principal components, denoted as \((\pi_1, \pi_2, \pi_3)\) in a Euclidean space. The principal components computed for each frame (i.e., sampling time instant) is represented by a circle marker and these points are joined by line segments. 
Overall, the geometric shape exhibited by the trajectories embodies a limit cycle, showing a good topological correspondence to the limit cycle as predicted by the model. In particular, from the view of Fig. \ref{fig:KKL}, a similar ``bow-tie'' shape can be seen. These results verify that the principal components, as ``reconstructed'' states, approximately form a diffeomorphism to the underlying unknown states. 
In addition, as can be observed in Fig. \ref{fig:Pixels}, there exists a steep stage in each period, namely when the red color component drops (and the other two curves also drop simultaneously); this is also seen in the model simulation in Fig. \ref{fig:Oregonator_x2} where an abrupt increase in \(x_2\) occurs in each period. In the PCA result (Fig. \ref{fig:KKL}), we indeed observe that there exists such long ``links'', on which the data points are scarce, between the two ends (low-\(\pi_1\) and high-\(\pi_1\) regions) with denser data. 
    
\par We note that the oscillations in Fig. \ref{fig:KKL} are in fact slowly decaying cycles instead of being fully periodic, which is consistent with the physical reality that this reaction system will in fact approach thermodynamic equilibrium, according to the second law, despite being slow. This decay is not captured by the simple Oregonator model, but honestly reflected by our proposed data-driven approach. 
Another apparent discrepancy between the PCA outcomes and the Oregonator model is that the principal components obtained are noisy and their trajectory is non-smooth. This naturally results from the non-ideal measurements of the camera (e.g., due to inaccurate recording of colors or slight deviation from the fixed position), as well as the intrinsic dynamic uncertainties (e.g., due to the incomplete mixing of the fluid in the reactive system).

\section{Conclusion}\label{sec:final}
\par To avoid the difficulty of model-based nonlinear state observer synthesis, this paper has considered a fully model-free, data-driven approach. Specifically, by exploiting the generic Kazantzis-Kravaris/Luenberger (KKL) observer structure, we propose to (i) use an a priori assigned LTI dynamics to convert the process outputs to observer states, which stand for a lifting of the states into a higher-dimensional space, and then (ii) perform a simple PCA algorithm to recover the underlying state dimensions. The resulting principal components provide a topologically equivalent representation of the true states, despite unmeasurable, through a diffeomorphism. 

\par The performance of the proposed approach relies on the assumption of a low-dimensional state space and high-dimensional outputs. This is the case for process units with dissipative dynamics, on which a camera captures real-time video measurements. In other words, \emph{when rich measurements are available, the data analysis required for obtaining the necessary state information can be largely or even surprisingly simplified}. 
Using the continuously stirred B-Z reaction system as a representative example, the proposed approach recovered its qualitative behavior, where oscillations are observed with clear topological correspondence to the limit cycle predicted by a standard model. 
As such, the present work makes a case for adopting image-based (and in general, non-traditional) sensing technology for industrial process control and monitoring purposes. 

\par The current work, however, is restricted to autonomous systems, where the LTI dynamics in the KKL observer can be a priori given. For non-autonomous, controlled systems, a KKL observer should theoretically involve a nonlinear state-dependent term in its dynamical part \citep{bernard2018luenberger}. The learning of such a data-driven KKL observer with more complex structure, is being actively studied. Also, on the basis of learned state observers, we will aim at achieving model-free feedback control in the future research. 

\bibliography{mybib}

\end{document}